# Energy-efficient PON-based Backhaul Connectivity for a VLC-enabled Indoor Fog Computing Environment

Wafaa B. M. Fadlelmula, Sanaa Hamid Mohamed, Taisir E. H. El-Gorashi, Jaafar M. H. Elmirghani

*Abstract*— In this paper, we consider the use of visible light communication (VLC) to provide connectivity to indoor fog computing resources and propose an energy-efficient passive optical network (PON)-based backhaul architecture to support the VLC system. We develop a mixed-integer linear programming (MILP) model to optimize the allocation of computing resources over the proposed architecture, aiming to minimize processing and networking power consumption. We evaluate the performance of the proposed architecture under varying workload demands and user distributions. Comparative analysis against a backhaul architecture that is based on the state-of-the-art spine-and-leaf (S&L) network design demonstrates total power savings of up to 82%. Further comparison with centralized cloud processing shows improvements in energy efficiency of up to 93%. Additionally, we examine the improvements in energy efficiency obtained by splitting tasks among multiple processing nodes and propose enhancements to the architecture including dynamic bandwidth allocation, increased wavelength bandwidth and improved connectivity within rooms to alleviate networking bottlenecks. Furthermore, we introduce an inter-building architecture that leverages resources from neighboring buildings to support high-demand scenarios.

## I. INTRODUCTION

The number of connected devices is growing exponentially and is projected to reach 500 billion by 2030; an enormous increase from 23 billion in 2018. This rapid growth is generating vast amounts of data, calling for innovative computing approaches to effectively process and manage data [1]. Fog computing has emerged as a computing paradigm that complements the centralized services of cloud computing by offering computational and storage resources in users' geographical proximity. The proximity of resources reduces the distances traversed by traffic flows, reducing latency and power consumption [6]. The delivery of efficient fog computing services requires careful design of the access network. Given that users spend 80%-90% of their time in indoor environments, with 80% of wireless data traffic is traced back to indoor users [3], special attention should be given to indoor access networks to provide energy efficient indoor connectivity.

As one of the enabling technologies for the sixth generation (6G) networks, visible light communication (VLC) emerged as a promising technology for indoor communication to provide energy efficient high data rate communication in conjunction with illumination [9]. VLC offers key advantages including license-free operation, electromagnetic immunity to interference, and enhanced security through physical confinement. Effective large-scale VLC deployments require effective backhauling to connect VLC access points indoor environment to the broader network. Several technologies were proposed to provide backhaul connectivity such as power line communication (PLC) [10], Ethernet over light (EoL) [11], optical fiber backhaul [12], passive optical networks (PONs) [13], and wireless backhauls [14]. Among these technologies, PONs can provide bidirectional, cost-effective, and high data rate communication while maintaining energy efficiency through the utilization of passive components. In [15], the authors proposed PON-based architectures to provide energy efficient interconnections for cloud and fog data centers (DCs) where both east-west communication (i.e., for local computations among multiple servers) and north-south communication (i.e., the connectivity with wider networks) are required.

In this paper, we propose a novel architecture that adopts wavelength routed PONs to provide wired backhaul connectivity for a VLC-enabled indoor fog computing environment. The architecture is designed to enable energy efficient networking for east-west as well as north-south traffic. To optimize resource allocation in this architecture, we develop a mixed integer linear programming (MILP) model that determines the optimal placement of computational tasks across fog nodes and cloud resources while minimizing the total power consumption of both processing and networking. We compare the energy efficiency of our proposed architecture against the use of state-of-the-art spine-and-leaf (S&L) architecture for backhauling. We evaluate the improvement in the energy efficiency of fog computing services over the proposed architecture compared to centralized cloud services. Also, we examine the improvements in energy efficiency obtained by splitting tasks among processing nodes and by utilizing dynamic bandwidth allocation (DBA), increased wavelength bandwidth and improved connectivity within rooms to improve the PON network. Furthermore, we examine an inter-building architecture where resources from neighboring buildings are utilized under high-demand scenarios.

The remainder of the paper is organized as follows. In Section II. The In-Building Network, we describe the system model of the proposed PON-based backhaul architecture along with a S&L-based backhaul architecture. Section III. MILP Models for Optimal Processing Demand Allocation presents the MILP models formulated to optimally allocate processing resources while minimizing power consumption. In Section V Design Enhancement to The Proposed PON Architecture, we present the evaluation scenarios and discuss the results of our optimization models. In Section0 V Design Enhancement to The Proposed PON Architecture, we explore enhancements to the proposed PON-based architecture to further improve energy efficiency. Finally, Section VI. Conclusion provides the conclusions and directions for future research.

## II. THE IN-BUILDING NETWORK ARCHITECTURES

### A. The PON Architecture

The proposed PON architecture comprises an in-building network, where a building of several rooms is considered, as shown in **Error! Reference source not found.**. Each room has several VLC access points serving multiple user devices. The suggested architectures are based on enhanced PONs that provide energy-efficient fabric [15]. The architecture is supported by fog nodes such as user mobile devices, and processing resources at the edge of the network. Also, processing resources are available at the building, the campus, the metro network, and at the cloud. The main components that constitute the system model are as follows:

*1. VLC Indoor System*

We consider a set of laser diodes (LDs) mounted on the ceiling of each room to act as light sources and transmitting units. To generate the visible white light, a combination of red, yellow, blue, and green is used [16]. WDMA is used to support multiple access where a multiplexer is attached to each transmitter to combine different wavelengths into an optical wireless beam. Using the red wavelength provides the highest data rate, but the lowest energy efficiency, while using the yellow wavelength offers lower data rates compared to the red wavelength but higher energy efficiency. Compared to the yellow wavelength, the green and blue wavelengths offer lower data rates and higher energy efficiency. At the user devices, an angle diversity receiver (ADR) is considered. By using a demultiplexer at each branch of the receiver, the received WDM optical beam is separated into multiple wavelengths. The optical beam reaching each branch of the receiver can be received as a line-of-sight (LOS) signal from the access point or as non-line-of-sight (NLOS) signal that is reflected by a different medium in the room. Each user device is configured to be wirelessly connected to a single access point using a single wavelength and a single branch of the ADR. VLC is mainly utilized to support downlink transmission in the indoor system. Adopting visible light for uplink is unpopular for portable devices due to the glare associated with VLC uplinks. Therefore, other alternatives can be adopted for uplink communication including RF, infrared, near UV and retro-reflective transceivers. However, to avoid complexity in this work, we assume that VLC is used also for the uplink communication.

*2. PON-based Backhaul Network*

The PON-based backhaul network provides passive connectivity between the distributed access points within the building. Access points and a room fog server (RFS) form a single PON group. The proposed design is composed of:

- Optical Network Units (ONUs): Each of the access points and RFSs are equipped with a tunable ONU that can be tuned to the respective wavelength based on communication requirements.

- Optical line terminal (OLT): The OLT aggregates the signals from access points and RFSs to be forwarded to the rest of the network. The OLT also demultiplexes signals in the opposite direction. The communication between the OLT and the PON groups is achieved via a hybrid TWDM PON [17]. In addition, the OLT controls and manages the PON and assigns wavelengths to the PON groups.

- Optical combiners/splitters: A single combiner is used in each room to combine upstream traffic from the several devices within each PON group into a single optical fiber while a single splitter is used to divide the downstream optical signal from a single fiber into several devices within each PON group.

- Arrayed Waveguide Gratings (AWGs): AWGs are used as $nx1$ multiplexers, where the AWG receives $n$ input signals with different wavelengths from the PON groups within the building and combines them into a single signal in the output port of the AWG that is connected to the OLT. Also, AWGs are used as $1xn$ demultiplexers, where a single signal is received at the input port of the AWG that is connected to the OLT and distributed to $n$ outputs.

- Arrayed Waveguide Grating Routers (AWGRs): AWGRs passively route the signals from input ports to output ports using a fixed routing matrix. Two ($5x5$) AWGRs are considered to facilitate passive transmission of the traffic in an indoor PON architecture with four rooms. Each splitter and coupler within a room are directly connected to the AWGRs. The couplers are attached to the input ports while the output ports are linked to the splitters. Two AWGs are used to connect the OLT to the AWGRs.

*3. The Fog Computing Architecture*

We consider a multi-layer fog architecture, where processing resources are placed at several locations at the edge of the network. The fog resources are summarized as follows.

- User Devices: We only consider a single type of mobile device in this work to avoid complexity. Each user device within each room can either generate job requests or offer their idle processing capabilities to requests from other user devices within the building. Nevertheless, the processing resources of the user devices are limited compared to other dedicated fog nodes.

- RFSs: Typical examples of such resources include personal laptops, or desktops inside the rooms. Homogenous laptops are assumed to be used as processing resource in each room.

- Building Fog Servers (BFSs): BFS can offer enhanced processing capacities relative to the RF layer. In contrast to RFSs, where personal devices can be used, in this layer, dedicated server(s) can be utilized.

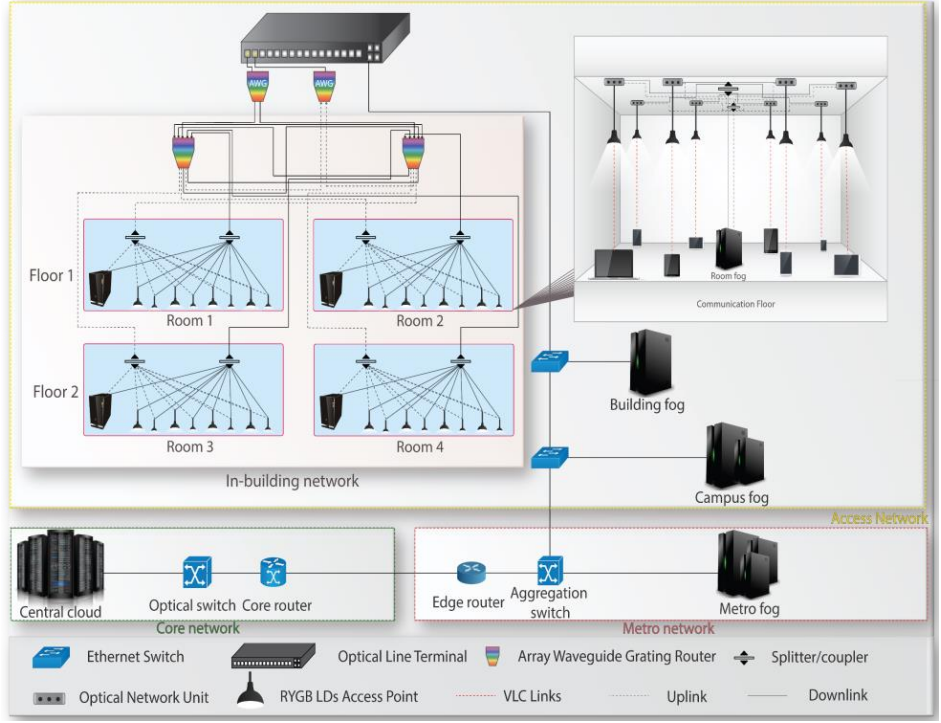

Fig. 1. The in-building PON-based VLC backhaul system with an end-to-end fog/cloud architecture.

- Campus Fog Servers (CFSs): The CFS can be a cluster of servers serving requests from multiple buildings within the campus. Note that CFS is located within the access network where an Ethernet switch is used to directly connect the CFS to the OLT. It is important to highlight that in this work, we only consider a single server in each layer to serve the demands.

- Metro Fog Servers (MFSs): The capacities of the fog resources in this layer are relatively high compared to the previous fog layers. This is due to the substantial number of users to be served by this node. Note that typically, the metro network consists of multiple routers which are connected to the core networks. However, we simplify the network into a single edge router and a single aggregation switch that is directly connected to the metro fog node.

- Central Cloud Servers (CCSs): Cloud DCs are located in the core network which aggregates the traffic from the metro networks. Cloud DC typically encompasses an extensive assemblage of servers to provide massive processing capabilities. We also assume using a single server from the CCSs to serve the tasks.

*B. Spine-and-Leaf-based Architecture*

The S&L architecture is proven to provide high bandwidth and low latency connectivity within DCs. Recently, S&L DCs have been adopted to support scalable and reliable fog computing architecture in the access network [18]. To leverage the S&L as a backhaul network architecture for a VLC-based system, the devices within each room are directly connected to one of the leaf switches, as shown Fig. 2. In this work, we consider two spine switches, where each spine is connected to all leaf switches and to a router. The router connects the access network to the rest of the network (the metro and core networks), through an Ethernet switch. Note that no OLT is used in the S&L backhaul system and hence, the building fog is connected to the other layers via switches only.

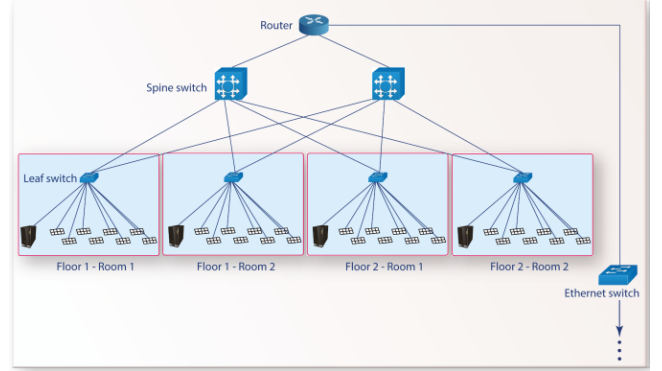

Fig. 2. The in-building S&L VLC backhaul system.

III. MILP MODELS FOR OPTIMAL PROCESSING DEMAND ALLOCATION

In this section, we present the MILP models we developed to optimally allocate the processing demands in the multi-layered fog architecture. The models minimize the processing and networking power consumption and consider the PON-based and the S&L-based backhaul architectures.

*A. PON-based Backhaul Architecture*

The sets, parameters, and variables of the MILP model for the PON-based backhaul architecture are defined as follows:

**Sets:**
$N$      Set of all nodes.
$N_i$      Set of neighbors of node $i$, $i \in N$.
$UD$      Set of user devices generating demands.
$PUD$      Set of user devices providing processing resources, $UD \cap PUD = \emptyset$.

| | |
|---|---|
| $RF$ | Set of RFSs. |
| $P$ | Set of processing nodes including the user devices, RFSs, BFS, CFS, MFS, and cloud server. |
| $AP$ | Set of access points connected to source nodes (i.e., user devices which generate requests). |
| $CP$ | Set of access points connected to processing user devices, $AP \cap CP = \emptyset$. |
| $MR$ | Set of user devices generating demands that use the red wavelength. |
| $MY$ | Set of user devices generating demands that use the yellow wavelength. |
| $MGB$ | Set of user devices generating demands that use green or blue wavelengths. |
| $DR$ | Set of processing user devices that use the red wavelength. |
| $DY$ | Set of processing user devices that use the yellow wavelength. |
| $DGB$ | Set of processing user devices that use the green or blue wavelengths. |
| $U$ | Set of ONUs attached to RFSs. |
| $Q$ | Set of networking devices including the OLT, Ethernet switches, aggregation switch, edge router, and the optical switch. |
| $G$ | Set of AWGRs. |
| $IN_g$ | Set of input ports of AWGR $g$, $g \in G$. |
| $OUT_g$ | Set of output ports of AWGR $g$, $g \in G$. |
| $W$ | Set of wavelengths for the PON. |

**Processing parameters:**

| | |
|---|---|
| $D_u$ | Processing demand generated by user device $u$, $u \in UD$. |
| $MP_d$ | Maximum power consumption of processing node $d$, $d \in P$. |
| $I_d$ | Idle power consumption of processing node $d$, $d \in P$. |
| $\Omega_d$ | Processing capacity of processing node $d$, $d \in P$. |
| $R_d$ | Power consumption per processing unit of processing node $d$, where $R_d = \left(\frac{MP_d - I_d}{\Omega_d}\right), d \in P$. |

**Networking parameters:**

| | |
|---|---|
| $R_u$ | Data rate demand generated by user device $u$, $u \in UD$. |
| $L_{ijw}$ | Capacity of physical link $(i,j)$, $i \in N, j \in N_i$, using wavelength $w$, $w \in W$. |
| $DRR_u$ | Data rate ratio, where $DRR_u = \left(\frac{R_u}{D_u}\right), u \in UD$. |
| $PR$ | Maximum power consumption of an access point when serving a user device using the red wavelength. |
| $IR$ | Idle power consumption of an access point when serving a user device using the red wavelength. |
| $BP$ | Bit rate of an access point. |
| $\Lambda$ | Power consumption per bit for an access point when serving a user device using the red wavelength, where $\Lambda = \frac{PR - IR}{BP}$. |
| $PY$ | Maximum power consumption of an access point when serving a user device using the yellow wavelength. |
| $IY$ | Idle power consumption of an access point when serving a user device using the yellow wavelength. |
| $\varsigma$ | Power consumption per bit for an access point when serving a user device using the yellow wavelength, where $\varsigma = \frac{PY - IY}{BP}$. |
| $BGP$ | Maximum power consumption of an access point when serving a user device using the green or blue wavelengths. |
| $IGP$ | Idle power consumption of an access point when serving a user device using green or blue wavelengths. |
| $\Pi$ | Power consumption per bit for an access point when serving a user device using the green or blue wavelengths, where $\Pi = \frac{PGB - IGB}{BP}$. |
| $PU$ | Maximum power consumption of an ONU. |
| $\mathbb{U}$ | Idle power consumption of an ONU. |
| $BU$ | Bit rate of an ONU. |
| $X$ | Power consumption per bit for an ONU, where $X = \frac{PU - \mathbb{U}}{BU}$. |
| $MQ_i$ | Maximum power consumption of networking device $i$, $i \in Q$. |
| $\gamma_i$ | Idle power consumption of networking device $i$, $i \in Q$. |
| $BQ_i$ | Bit rate of networking device $i$, $i \in Q$. |
| $Ч_i$ | Power consumption per bit for networking device $i$, $i \in Q$, where $Ч_i = \frac{MQ_i - \gamma_i}{BQ_i}$. |
| $Z$ | A very large number. |

**Variables:**

| | |
|---|---|
| $\lambda_{ijw}^{ud}$ | Traffic flow between source and destination pair $(u,d)$ traversing the physical link $(i,j)$ using wavelength $w$, where $u \in UD$, $d \in P, i \in N, j \in N_i, w \in W$. |
| $\lambda_w^{ud}$ | Traffic flow between source and destination pair $(u,d)$ using wavelength $w$, where $u \in UD, d \in P, w \in W$. |
| $w_{ij}^{ud}$ | Volume of traffic between source and destination pair $(u,d)$ on the physical link $(i,j)$, where $u \in UD, d \in P, i \in Q, j \in N_i$. |
| $\lambda_i$ | Volume of traffic aggregated by node $i$, where $i \in Q$. |
| $\sigma_i$ | Volume of traffic received by or aggregated by node $i$, where $i \in CP \cup PUD$. |
| $\mu_i$ | Volume of traffic generated or aggregated by node $i$, where $i \in AP \cup UD$. |
| $\psi^{ud}$ | Processing demand generated by source node $u$ and served by a processing node $d$, where $u \in UD, d \in P$. |
| $\Theta_i$ | Binary variable, $\Theta_i = 1$, if networking node $i$ is activated, otherwise, $\Theta_i = 0$, where $i \in Q$. |
| $\xi^{ud}$ | Binary variable, $\xi^{sd} = 1$, if the processing demand generated by user device $u$ is assigned to processing node $d$, otherwise, $\xi^{ud} = 0$, where $u \in UD, d \in P$. |
| $\beta_d$ | Binary variable, $\beta_d = 1$, if processing node $d$ is activated, otherwise, $\beta_d = 0$, where $d \in P$. |
| $\rho_w^{ud}$ | Binary variable, $\rho_w^{ud} = 1$, if wavelength $w$ is assigned to carry the traffic between source and destination pair $(u,d)$, otherwise $\rho_w^{ud} = 0$, where $u \in UD, d \in P, w \in W$. |
| $\alpha, \tau$ | Weight factors used to determine the significance of the decision variables. |

The variables are related as follows:

$$\mu_i = \sum_{u \in UD} \sum_{d \in P, u \neq d} \sum_{j \in N_i} \sum_{w \in W} \lambda_{ijw}^{ud}$$
$$+ \sum_{u \in UD} \sum_{d \in P, u \neq d} \sum_{j \in N_i} \sum_{w \in W} \lambda_{jiw}^{ud}, \quad (1)$$
$$\forall i \in AP \cup UD.$$

Equation

(1) calculates the total traffic volume $\mu_i$ handled by node $i$, where $i$ can be either a user device ($i \in UD$) generating traffic or an access point ($i \in AP$) aggregating traffic from connected source user devices.

$$\sigma_i = \sum_{u \in UD} \sum_{d \in PUD \cup RF} \sum_{j \in N_i, i \neq u} \sum_{w \in W} \lambda_{ijw}^{ud}$$
$$+ \sum_{u \in UD} \sum_{d \in PUD \cup RF} \sum_{j \in N_i, i \neq u} \sum_{w \in W} \lambda_{jiw}^{ud}, \quad (2)$$
$$\forall i \in CP \cup PUD.$$

Equation

(2) calculates the total traffic aggregated by an access point connected to a user device providing processing or an ONU connected to the RFS. The traffic comprises the traffic within the rooms due to processing the demands in idle user devices offering processing or the RFSs.

$$\lambda_i = \sum_{u \in UD} \sum_{d \in P, u \neq d} \sum_{j \in N_i, i \neq u} w_{ij}^{ud}$$
$$+ \sum_{u \in UD} \sum_{d \in P, u \neq d} \sum_{j \in N_i, i \neq u} w_{ji}^{ud}, \quad (3)$$
$$\forall i \in Q.$$

Equation

(3) calculates the total traffic aggregated by the networking devices outside the building including the OLT, Ethernet switches, aggregation switch, the edge router, the core router, and the optical switch. This traffic comprises the traffic that leaves the room due to processing at the BFS, CFS, MFS, or the cloud.

The objective of the model is to minimize the total power consumption, while maximizing the number of accepted processing tasks:

$$\text{Minimize} \quad \alpha \, TPC - \tau \, Y. \quad (4)$$

where $TPC$ is the total power consumption and $Y$ is the total number of tasks assigned for processing nodes. $TPC$ is composed of both processing power consumption ($PC$) and networking power consumption ($PN$):

$$TPC = PC + PN. \quad (5)$$

The processing power consumption ($PC$) is given as:

$$PC = \sum_{u \in UD} \sum_{d \in P} \psi^{ud} \, R_d + \beta_d \, I_d. \quad (6)$$

The networking power consumption ($PN$) is given as:

$$PN = PAP + PCP + AR + AY + AGB + PR$$
$$+ PY + PGB + PRS + PQ. \quad (7)$$

where:

$PAP$ is the power consumption of ONUs attached to access points connecting source user devices, given as:

$$PAP = \sum_{i \in AP} \mu_i X + \Theta_i \mathbb{U}. \quad (8)$$

$PCP$ is the power consumption of ONUs attached to access points connecting processing user devices, defined as:

$$PCP = \sum_{i \in CP} \sigma_i X + \Theta_i \mathbb{U}. \quad (9)$$

$AR, AY,$ and $AGB$ are the power consumption of access points connected to source user devices using red, yellow, and green/blue wavelengths, respectively. These are defined as:

$$AR = \sum_{i \in MR} \mu_i \Lambda + \Theta_i IR. \quad (10)$$

$$AY = \sum_{i \in MY} \mu_i \lambda + \Theta_i IY. \quad (11)$$

$$AGB = \sum_{i \in MGB} \mu_i \Pi + \Theta_i IGB. \quad (12)$$

$PR, PY,$ and $PGB$ are the power consumption of access points connected to processing user devices using red, yellow, and green/blue wavelengths, respectively. These are expressed as:

$$PR = \sum_{i \in DR} \sigma_i \Lambda + \Theta_i IR. \quad (13)$$

$$PY = \sum_{i \in DY} \sigma_i \lambda + \Theta_i IY. \quad (14)$$

$$PGB = \sum_{i \in DGB} \sigma_i \Pi + \Theta_i IGB. \quad (15)$$

$PRS$ is the power consumption of ONUs attached to RFSs, given as:

$$PORS = \sum_{i \in U} \sigma_i X + \Theta_i \mathbb{U}. \quad (16)$$

$PQ$ is the power consumption of the networking devices, defined as:

$$PQ = \sum_{i \in Q} \lambda_i \, \mathtt{Ч}_i + \Theta_i \gamma_i. \quad (17)$$

The total number of tasks assigned to processing nodes ($Y$) is given as:

$$Y = \sum_{u \in UD} \sum_{d \in P} \xi^{ud}. \quad (18)$$

The model is subject to the following constraints:

$$\psi^{ud} = D_u \xi^{ud}, \forall u \in UD, d \in P. \quad (19)$$

Constraint (19) indicate that the processing demand of user device $u$ is either accepted and fully assigned to a processing node $d$, or rejected.

$$\sum_{u \in UD} \psi^{ud} \leq \Omega_d, \forall d \in P. \quad (20)$$

Constraint (20) ensures that the total processing demands allocated to a processing node do not exceed its capacity.

$$\sum_{w \in W} \lambda_w^{ud} = DRR_u \, \psi^{ud}, \forall u \in UD, d \in P. \quad (21)$$

Constraint (21) ensures that the traffic flow between a source and destination pair over all wavelengths is equal to the traffic demand.

$$\sum_{j \in N_i, i \neq j} \lambda_{ijw}^{ud} - \sum_{j \in N_i, i \neq j} \lambda_{jiw}^{ud} = \begin{cases} \lambda_w^{ud} & i = u \\ -\lambda_w^{ud} & i = d \\ 0 & \text{otherwise,} \end{cases} \quad (22)$$
$$\forall u \in UD, d \in P, i \in N, w \in W.$$

Constraint (22) is the flow conservation constraint to ensure that the total incoming traffic at a node at a certain wavelength is equal to the outgoing traffic at the same wavelength except for the source and destination nodes.

$$\sum_{u \in UD} \sum_{d \in P} \lambda_{ijw}^{ud} \leq L_{ijw}, \quad (23)$$
$$\forall i \in N, j \in N_i, w \in W, u \neq d.$$

Constraint (23) ensures that the total traffic flow between source and destination pair $(u,d)$ carried on link $(i,j)$ using given wavelength $w$, does not exceed the capacity of that wavelength.

$$Z \xi^{ud} \geq \psi^{ud}, \forall u \in UD, d \in P. \quad (24)$$
$$\xi^{ud} \leq Z \psi^{ud}, \forall u \in UD, d \in P. \quad (25)$$

Constraints (24) and (25) are used to ensure that the binary variable $\xi_{ud}$ is related to the non-binary variable $\psi^{ud}$.

$$Z \beta_d \geq \sum_{u \in UD} \xi^{ud}, \forall d \in P. \quad (26)$$
$$\beta_d \leq Z \sum_{u \in UD} \xi^{ud}, \forall d \in P. \quad (27)$$

Constraints (26) and (27) are used to ensure that the binary variable $\beta_d$ related to the sum of the non-binary variable $\xi^{ud}$.

$$Z \Theta_i \geq \lambda_i, \forall i \in Q. \quad (28)$$
$$\Theta_i \leq Z \lambda_i, \forall i \in Q. \quad (29)$$

Constraints (28) and (29) ensure that the binary variable $\Theta_i$ is related to the non-binary variable $\lambda_i$.

$$Z \Theta_i \geq \sigma_i, \forall i \in CP \cup PUD. \quad (30)$$
$$\Theta_i \leq Z \sigma_i, \forall i \in CP \cup PUD. \quad (31)$$

Constraints (30) and (31) ensure that the binary variable $\Theta_i$ is related to the non-binary variable $\sigma_i$ for intermediate nodes of traffic flows.

$$Z \Theta_i \geq \mu_i, \forall i \in AP \cup UD. \quad (32)$$
$$\Theta_i \leq Z \mu_i, \forall i \in AP \cup UD. \quad (33)$$

Constraints (32) and (33) ensure that the binary variable $\Theta_i$ is related to the non-binary variable $\mu_i$ for access points aggregating traffic from user devices.

$$\sum_{d \in P} \xi^{ud} \leq 1, \forall u \in UD. \quad (34)$$

Constraint (34) ensures that each processing demand can be assigned to at most one processing node, allowing for the possibility of rejecting some demands.

$$Z \rho_w^{ud} \geq \lambda_w^{ud}, \forall u \in UD, d \in P, w \in W. \quad (35)$$
$$\rho_w^{ud} \leq Z \lambda_w^{ud}, \forall u \in UD, d \in P, w \in W. \quad (36)$$

Constraints (35) and (36) ensure that the binary variable $\rho_w^{ud}$ is is related to the non-binary variable $\lambda_w^{ud}$.

$$\sum_{w \in W} \rho_w^{ud} \leq 1, \forall u \in UD, d \in P. \quad (37)$$

Constraint (37) ensures that no more than one wavelength is assigned for a demand.

$$\sum_{w \in W} \lambda_{ijw}^{ud} = w_{ij}^{ud}, \quad (38)$$
$$\forall u \in UD, d \in P, i \in Q, j \in N_i.$$

Constraint (38) aggregates the traffic from all wavelengths to calculate the traffic going through the OLT, the metro network devices, and the core network devices.

$$\sum_{j \in IN_G} \lambda_{ijw}^{ud} \leq 0, \quad (39)$$
$$\forall u \in UD, d \in P, g \in G, i \in OUT_g, w \in W.$$

Constraint (39) enforces the traffic flow to be directed from the input ports to the output ports of the AWGRs.

*B. Spine-and-leaf-based backhaul System Model*

The MILP model in Section III.A is modified to represent the S&L-based architecture. All the parameters, variables, and constraints related to the PON-based architecture are omitted and additional sets, parameters, and variables are added to the model. The sets of the model are all maintained except $G$, $IN_g$, $OUT_g$, $W$ and $U$. The parameters $PU$, $\mathbb{U}$, $BU$ and $X$ related to the ONU are removed. The parameter $L_{ijw}$ is replaced by $L_{ij}$, as conventional S&L DCs do not use WDM. The sets and parameters that define the topology such as $N$ and $N_i$ are updated to reflect the S&L architecture. Variables $W_{ij}^{ud}$, $\rho_w^{ud}$ are removed and variables $\lambda_w^{ud}$, $\lambda_{ijw}^{ud}$ are replaced by $\lambda^{ud}$, $\lambda_{ij}^{ud}$ which are defined as new variables.

The following are the new sets, parameters, and variables.

**Sets:**
$SL$     Set of networking devices related to the S&L architecture including the leaf switch, spine switch, and a router.

**Parameters:**
$L_{ij}$     Capacity of the physical link $(i,j), i \in N, j \in N_i$.
$MS_i$     Maximum power consumption of networking device $i, i \in SL$.
$BS_i$     Bit rate of networking device $i, i \in SL$.
$F_i$     Power consumption per bit of network device $i$, where $F_i = \frac{MS_i - \varsigma_i}{BS_i}, i \in SL$.
$T$     Maximum power consumption of an Ethernet interface attached to an access point.
$C$     Idle power consumption of an Ethernet interface attached to an access point.
$BI$     Bit rate of the Ethernet interface.
$\flat$     Power consumption per bit of the Ethernet interface, where $\flat = \frac{T-C}{BI}$.

**Variables:**
$PL$     Power consumption of the networking devices of the S&L architecture.
$\lambda^{ud}$     Traffic flow between source and destination pair $(u,d), u \in UD, d \in P$.

Equations

(*1*),

(2), and (3) are replaced by Equations (40), (41), and (42), respectively as follows:

$$\mu_i = \sum_{u\in UD}\sum_{d\in P, u\neq d}\sum_{j\in N_i}\lambda_{ij}^{ud} + \sum_{u\in UD}\sum_{d\in P, u\neq d}\sum_{j\in N_i}\lambda_{ji}^{ud}, \quad (40)$$
$$\forall i \in AP \cup UD.$$

$$\sigma_i = \sum_{u\in UD}\sum_{d\in PUD\cup RF}\sum_{j\in N_i, i\neq u}\lambda_{ij}^{ud}$$
$$+ \sum_{u\in UD}\sum_{d\in PUD\cup RF}\sum_{j\in N_i}\lambda_{ji}^{ud}, \quad (41)$$
$$\forall i \in CP \cup PUD.$$

$$\lambda_i = \sum_{u\in UD}\sum_{d\in P, u\neq d}\sum_{j\in N_i, i\neq u}\lambda_{ij}^{ud}$$
$$+ \sum_{u\in UD}\sum_{d\in P, u\neq d}\sum_{j\in N_i, i\neq u}\lambda_{ji}^{ud}, \quad (42)$$
$$\forall i \in Q.$$

The objective is expressed by Equation (4), where $PC$, the power consumption of computing is calculated using Equation (6), and $PN$, the total networking power consumption, is calculated by the following Equation:
$$PN = IAP + ICP + AR + AY + AGB + PR + PY + PGB + PQ + PL, \quad (43)$$
where, $AR, AY, AGB, PR, PY, PGB$, and $PQ$ are calculated by Equations, (10), (11), (12), (13), (14), (15) and (17), respectively. $PL$, the power consumption of the networking devices of the S&L architecture (i.e., the leaf switch, the spine switch, the router), is calculated as:
$$PL = \sum_{i \in SL}\lambda_i f_i + \theta_i \zeta_i. \quad (44)$$

$IAP$, the power consumption of the Ethernet interfaces attached to the access points connecting source user devices, is calculated as:
$$IAP = \sum_{i \in AP}\mu_i b + \theta_i C. \quad (45)$$

Note that typically an Ethernet interface is attached to servers. Therefore, no additional power consumption is considered for RFSs.

$ICP$, the power consumption of the Ethernet interfaces attached to the access points connecting processing user devices, is calculated as:
$$ICP = \sum_{i \in CP}\sigma_i b + \theta_i C. \quad (46)$$

The model is subject to the following constraints:
Constraints (19) to (21), (24) to (30), and (32) to (34) from Section III.B.
Equation (22) is replaced by Equation (47).

$$\sum_{j\in N_i, i\neq j}\lambda_{ji}^{ud} - \sum_{j\in N_i, i\neq j}\lambda_{ji}^{ud} = \begin{cases}\lambda^{ud} & i = u \\ -\lambda^{ud} & i = d \\ 0 & \text{otherwise,}\end{cases} \quad (47)$$
$$\forall u \in UD, d \in P, i \in N.$$

Constraint (47) is the flow conservation constraint which ensures the total incoming traffic at a node is equal to the outgoing traffic except for the source and destination nodes.
Equation (23) is replaced by Equation (48).
$$\sum_{u\in UD}\sum_{d\in P}\lambda_{ji}^{ud} \leq L_{ij}, \forall i \in N, j \in N_i, u \neq d. \quad (48)$$
Constraint (48) ensures that the total traffic flow between source and destination pair $(u,d)$ carried on link $(i,j)$ does not exceed the capacity of that link.
Equation (31) is replaced by Equation (49).
$$\lambda^{ud} = DRR_u \psi_{ud}, \quad (49)$$
$$\forall u \in UD, d \in P.$$
Constraint (49) ensures that the traffic flow between a source and destination pair is equal to the total traffic demand.

## IV. PERFORMANCE EVALUATION

In this section, we compare the performance of the PON architecture to the S&L architecture by examining three scenarios with varying users demands. These scenarios differ in the number of user devices generating tasks in each room, and hence, differ in terms of the total demands to be served. All the scenarios are examined under the same building structure where each room has eight VLC access points, with each access point serving either a single or multiple users based on proximity of users to the access points and the signal-to-interference-plus-noise ratio (SINR), following the approach described in [16]. According to [16], the red, yellow, green, and blue wavelengths used in the VLC system provide data rates of 2.5 Gbps, 2.5 Gbps, 2.5 Gbps, and 2.25 Gbps, respectively. In each scenario, we investigate the impact of user locations within the room by considering two cases. In the first case, all users are in the vicinity of two access points, utilizing the four wavelengths (i.e., red, yellow, green, and blue). In the second case, users are distributed across the room, where each user is located close to and served by a single access point. As high data rates are favored, each user in the distributed case is connected to a single access point using the red wavelength. Each user generates a single task, and we assume here that tasks are not splittable. The traffic demands are related to the processing demands by a data rate ratio (DRR) defined as the ratio of traffic demands in Gbps to processing demands in GFLOPs. We consider applications that are computation and communication intensive, such as computer vision applications including object detection and image classification. Such applications typically have a DRR of 0.05 with traffic demands in the range of 0.3 - 1 Gbps and processing workload demands per task in the range of 6-20 GFLOPs. Table 1 and Table 2 summarize the parameters of networking devices and processing nodes, respectively. Table 3 presents the S&L parameters [19].

TABLE 1: NETWORKING DEVICES PARAMETERS.

| Network device | Maximum power consumption (Watts) | Idle power consumption (Watts) | Capacity (GFLOPs) | Efficiency (Watts/ GFLOPs) |
|---|---|---|---|---|
| Red | 7.2 | 4.32 | 2.5 | 1.52 |
| Yellow | 4.5 | 2.7 | 2.5 | 0.72 |

| | | | | | |
|---|---|---|---|---|---|
| AP | Green | 2.7 | 1.62 | 2.5 | 0.432 |
| | Blue | 2.7 | 1.62 | 2.25 | 0.485 |
| ONU | | 15 | 9 | 10 | 0.9 |
| Ethernet switch | | 300 | 180 | 160 | 1.125 |
| Aggregation switch | | 435 | 261 | 240 | 0.725 |
| Edge router | | 435 | 261 | 240 | 0.725 |
| Optical switch | | 750 | 450 | 480 | 0.625 |
| Core router | | 344 | 206.4 | 3200 | 0.043 |

TABLE 2: PROCESSING DEVICES PARAMETERS.

| Processing node | Maximum power consumption (Watts) | Idle power consumption (Watts) | Capacity (GFLOPs) | Efficiency (Watts/ GFLOPs) |
|---|---|---|---|---|
| Cloud server | 1100 | 660 | 1612.8 | 0.27 |
| MFS | 750 | 450 | 403.2 | 0.74 |
| CFS | 350 | 210 | 121.6 | 1.15 |
| BFS | 305 | 183 | 99 | 1.23 |
| RFS | 65 | 39 | 64 | 0.41 |
| User devices | 18 | 10.8 | 12.288 | 0.55 |

TABLE 3: SPINE-AND-LEAF PARAMETERS.

| Processing node | Maximum power consumption (Watts) | Idle power consumption (Watts) | Capacity (GFLOPs) | Efficiency (Watts/ GFLOPs) |
|---|---|---|---|---|
| Leaf switch | 508 | 304.8 | 480 | 0.423 |
| Spine switch | 660 | 396 | 1440 | 0.183 |
| Router | 344 | 206.4 | 3200 | 0.043 |
| Ethernet interface | 5 | 3 | 10 | 0.2 |

We leverage a linear power profile to compute the power consumption of both processing and networking devices. This profile consists of idle and proportional parts. The idle part represents a constant amount of power ($P\_idle$) consumed as soon as the device is activated, regardless of the traffic load ($\lambda$) or processing demand. The proportional part scales linearly with the load, where maximum power ($P\_max$) is consumed when the device operates at full capacity ($C$). We assume that the idle part accounts for 60% of the maximum power consumption for all networking and the processing devices [20].

In this work, we only consider uplink traffic resulting from sending data associated with tasks requested by the users to their assigned processing resources. This is applicable for applications such as uploading video streams of high quality for processing. The downlink traffic, which may include acknowledgements of received content or instructions to room actuators, typically requires significantly lower data rates and is therefore not considered in our optimization model.

*A. Scenario 1: Low Intensity User Distribution*

In this scenario, only two user devices in each room generate processing demands while the remaining six idle user devices are available to process demands. Two cases representing different user locations are examined.

*1) Case 1: Clustered Users*

In this case, users in each room are in proximity to two access points. As observed in Fig. 3, the results show the optimal processing placement for both PON architecture (PON) and S&L architecture. In both architectures, all the demands are exclusively served within the rooms without the need to activate further remote fog units in the access or metro networks. The key difference between the two architectures lies in their resource selection strategy. In the PON architecture, the passive nature of connectivity enables uniform networking power consumption regardless of which nodes process the demands within the building. Therefore, the selection of processing resources depends primarily on the processing power consumption. Consequently, the PON architecture favors consolidating demands in RFSs when possible. For example, at lower demands (6-8 GFLOPs), a single RFS handles all processing. As the demands increase (9-12 GFLOPs), the capacity of the RFS is exhausted, and hence user devices are utilized to serve the demands. Even though RFSs are more efficient in terms of processing, their idle power contribution is considerably higher than the user devices. When the demands reach 13 GFLOPs, user devices are no longer a viable option due to their processing capacity limit, causing the model to activate additional RFSs instead. In contrast, the S&L architecture demonstrates a different behavior. Despite RFSs having higher processing efficiency, utilizing a single RFS requires traversing multiple switch levels, which significantly increases networking power consumption. Therefore, the S&L prioritizes local processing by idle user devices up to 12 GFLOPs. When demands exceed user devices capacity at 13 GFLOPs, the S&L architecture activates all four RFSs simultaneously to minimize the networking overhead associated with consolidated processing.

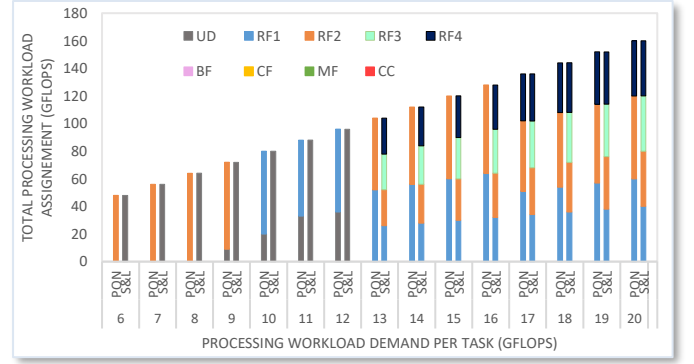

Fig. 3. Optimal processing placement in the PON and the S&L architecture for scenario 1- case 1.

Fig. 4 shows the power consumption of the PON and S&L architectures. The PON architecture is notably more efficient because of enabling the consolidation of demands into fewer processing nodes. Between 6-12 GFLOPs, both architectures show similar power consumption as four of the users in the same room share a single access point to relay the traffic without activating further networking nodes. Differences appear beyond 12 GFLOPs, when user devices processing capabilities are exhausted. The utilization of all RFSs in the S&L architecture leads to the utilization of leaf switches to route traffic within rooms which significantly increases the networking power consumption.

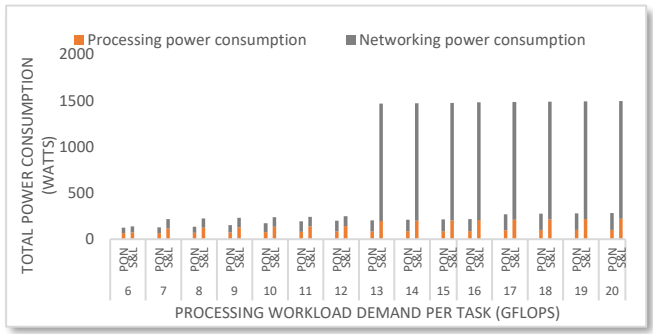

Fig. 4. Power consumption comparison for scenario 1 – case 1.

*2) Case 2: Distributed User*

In this case, users are scattered over the room and each access point serves a single user device using the red wavelength. Fig. 5 depicts the optimal processing placement in both architectures. For the PON architecture, for low demands (6-9 GFLOPs), the placement mirrors Case 1. At 10-12 GFLOPs, user devices are no longer a preferred option due to the high networking power cost of accessing multiple distributed user devices. Instead, two RFSs handle the demands until 17 GFLOPs, after which a third RFS becomes necessary. The S&L architecture shows similar placement to the first case at 6 GFLOPs only. From 7 GFLOPs onward, all RFSs are activated to accommodate demands. This shift occurs because serving distributed users require activating an access point for each user, whereas in Case 1 only two access points were needed.

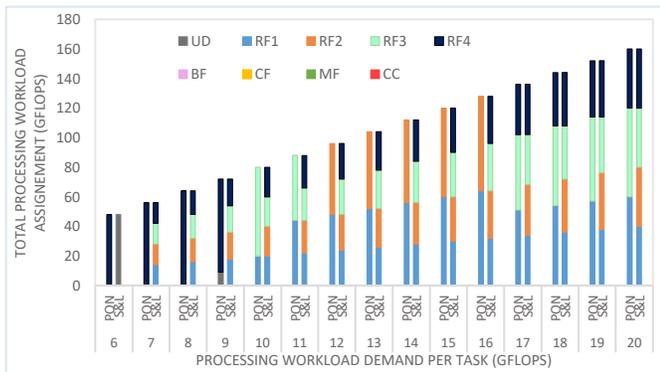

Fig. 5. Optimal processing placement in the PON architecture and the S&L architecture for scenario 1- case 2.

Fig. 6 presents the power consumption of the PON and S&L architectures showing that the PON architecture provides lower processing power consumption compared to S&L due to the activation of fewer processing nodes. Additionally, user distribution negatively affects the networking power consumption of S&L because of the activation of multiple leaf switches.

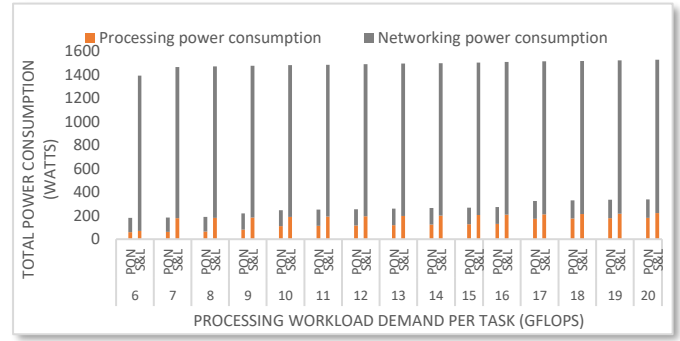

Fig. 6. Power consumption comparison for scenario 1 – case 2.

*B. Scenario 2: High Intensity User Distribution*

In this scenario, we examine the effect of increasing user demands in each room on the optimal processing placement. Each room has six user devices generating demands, with two remaining serving as processing nodes. Similar cases to those of Scenario 1 are examined.

*1) Case 1: Clustered Users*

In this case, the users are grouped into two groups of four user devices per group, each served by a single access point. As illustrated in Fig. 7, both architectures initially utilize local resources before activating fog nodes outside the building. At lower demands (6-9 GFLOPs), the PON architecture uses the fewest possible nodes (RFSs and user devices). As the demands increase, all RFSs are activated. At 11-12 GFLOPs, user devices are activated to handle the demands. However, despite RFSs having sufficient processing capacity for additional tasks, network constraints force the activation of eight user devices to serve tasks. As the demands increase (13-15 GFLOPs), the CFS is employed to serve the demands in addition to the RFSs. Note that the BFS is not used due to its limited capacity. From 16 GFLOPs onwards, the BFS is activated to handle the remaining demands. In the S&L, processing demands are served locally within the building at lower demand levels (6-12 GFLOPs) to avoid networking costs associated with routing through the spine switches. At 13 GFLOPs, the S&L activates the CFS. The BFS is activated at 15 and 16 GFLOPs. A key difference between the two architectures appears from 18 GFLOPs, where the S&L architecture selects the MFS along with RFSs as optimal locations for processing, while the PON architecture encounters network limitations and rejects some tasks.

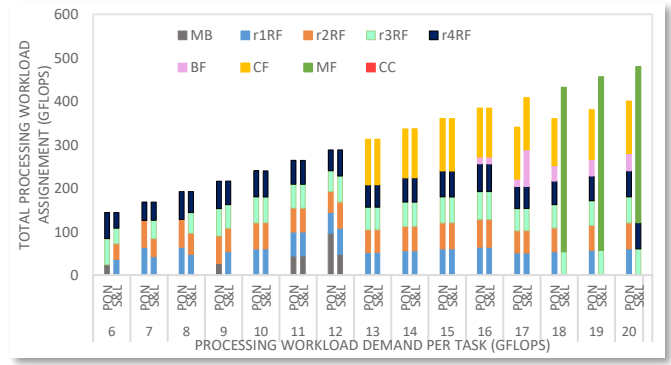

Fig. 7. Optimal processing placement in the PON architecture and the S&L architecture for scenario 2 - case 1.

Fig. 8. present the power consumption of the PON and the S&L architectures for demands ranging from 6-16 GFLOPs. In the PON architecture, the processing power consumption is lower for 6-9 GFLOPs due to the use of fewer processing nodes. For the remaining demands, both architectures exhibit similar processing power consumption, with an exception at 12 GFLOPs where the PON architecture experiences a network bottleneck that prevents efficient resource utilization. However, networking power consumption is substantially higher in the S&L architecture throughout the range. Additionally, a notable spike in networking power consumption occurs in both architectures from 13 GFLOPs onward due to the activation of remote processing nodes.

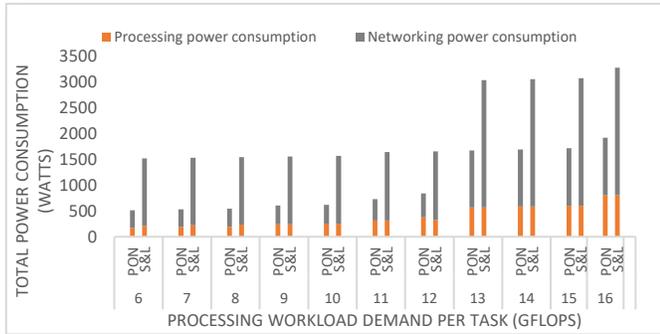

Fig. 8. Power consumption comparison for scenario 2 – case 2.

*3) Case 2: Distributed Users*

The second case examines distributed users within each room. Fig. 9 depicts the optimal processing placement for both architectures. In the PON architecture, network capacity constraints were not encountered since each user connects to a dedicated access point with full wavelength capacity which results in no task rejection. A notable difference from the first case occurs at 9 GFLOPs, where four RFSs are selected to host demands instead of user devices, primarily due to the network costs. From 18 GFLOPs onwards, the PON architecture activates MFS to handle demands along with RFSs. The S&L architecture experiences no change in the allocation of all the demands compared to Case 2.

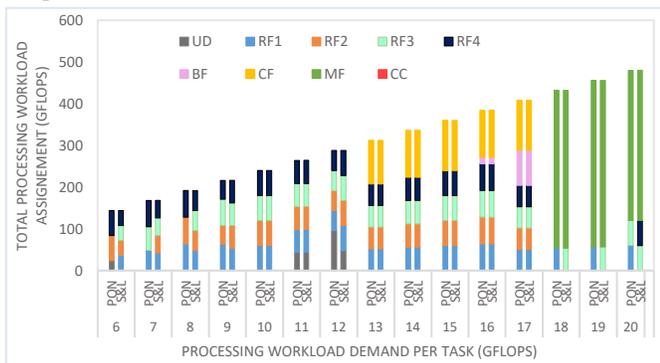

Fig. 9. Optimal processing placement in the PON architecture and the S&L architecture for scenario 2 - case 2.

Fig. 10 shows that power consumption in this case largely mirrors Case 2 with some exceptions. At 12 GFLOPs, the absence of network capacity constraints enables the PON architecture to achieve higher processing resource utilization which leads to improved processing power consumption. Additionally, both architectures demonstrate comparable processing power consumption values in the 17-20 GFLOPs range. The networking power consumption patterns continue to follow trends established in previous demands.

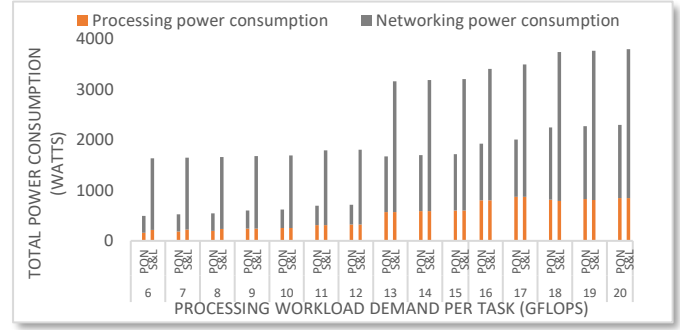

Fig. 10. Power consumption comparison for scenario 2 – case 2.

*C. Scenario 3: Demands from A Single Room*

In this scenario, eight user devices in a single room generate demands while the remaining user devices across all rooms offer their processing resources.

*1) Case 1: Clustered Users*

Fig. 11**Error! Reference source not found.** illustrates the processing placement in both architectures under Scenario 3 with users organized into two clusters of four devices, each served by a single access point. The placement differs significantly from previous scenarios. In the PON architecture, unlike the previous scenarios where RFSs were preferred, user devices are selected to host the demands from 6 to12 GFLOPs. This occurs because, from each room, each RFS can only be accessed through one wavelength, creating a bottleneck when multiple source devices within the same room try to access the same RFS simultaneously. When several clustered user devices attempt to transmit data through the same access using the same wavelength, they exceed the capacity limit of that wavelength. At 7 GFLOPs, despite having sufficient capacity in a single RFS, two RFSs are activated due to the wavelength limitations. Beyond 13 GFLOPs, the PON architecture fails to place all demands due to the network constraints and rejects one task while activating all RFSs and BFS. Placement in the S&L architecture follows a different approach in this scenario. Demands are allocated to the RFS in the same room, as all user devices can access it using a single leaf switch. At 9-10 GFLOPs, user devices from different rooms supplement the RFS using a single spine switch to convey traffic between rooms. As the demands increase, additional RFSs are utilized to serve the demands, Unlike the PON architecture, the S&L architecture successfully accommodates all tasks without rejection.

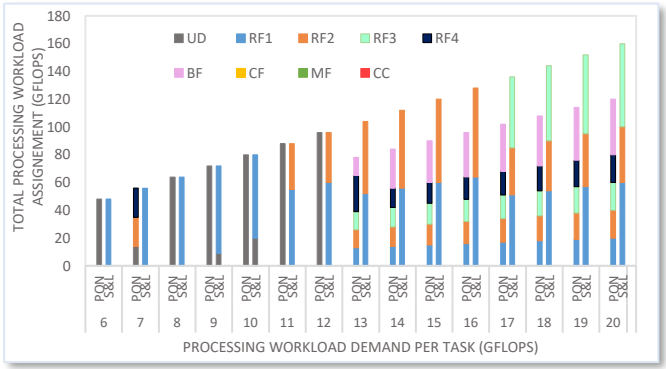

Fig. 11. Optimal processing placement in the PON architecture and the S&L architecture for scenario 3 – case 1.

Fig. 12 compares the power consumption comparison of the two architectures for processing demands ranging from 6-12 GFLOPs. The results show that S&L architecture consumes significantly less power for processing workloads. Conversely, the PON architecture maintains significantly greater efficiency in networking power consumption. Notably, the S&L architecture experiences a substantial increase in networking power consumption from 9 GFLOPs onwards due to the activation of spine switch to route traffic between rooms.

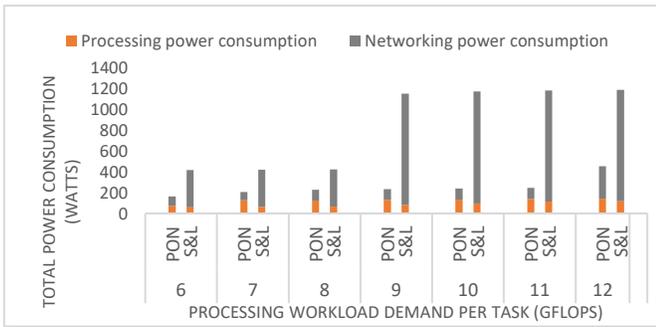

Fig. 12. Power consumption comparison for scenario 3 – case 1.

*2) Case 2: Distributed Users*

Fig. 13 illustrates the resource placement when users are distributed across the room. The PON architecture demonstrates notable differences compared to Case 1. RFS is selected instead of user devices at 7 GFLOPs due to the network cost associated with accessing other user devices. Network capacity constraints limit the full utilization of RFSs, so tasks are distributed across all RFSs from 8-11 GFLOPs. From 13 GFLOPs onwards, despite the higher power consumption, the BFS is activated alongside RFSs. The placement in the S&L architecture is consistent with Case 1.

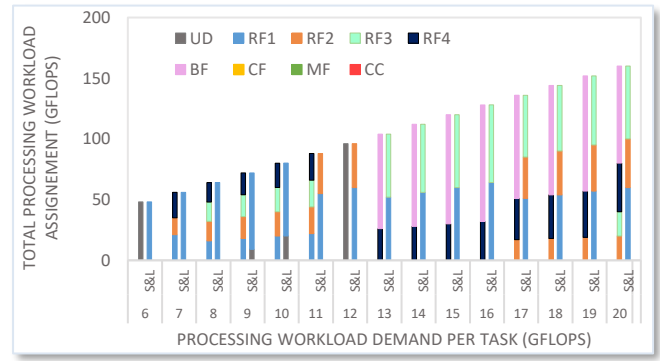

Fig. 13. Optimal processing placement in the PON architecture and the S&L architecture for scenario 3 – case 2.

Fig. 14 compares the power consumption of the PON and the S&L architectures. The results further emphasize the trends observed in Case 1. The networking power consumption of the PON architecture increases from 13 GFLOPs onwards due to its inability to utilize in-building resources. Additionally, the S&L architecture exhibits higher network power consumption beyond 17 GFLOPs when additional processing nodes are activated due to the utilization of multiple switches.

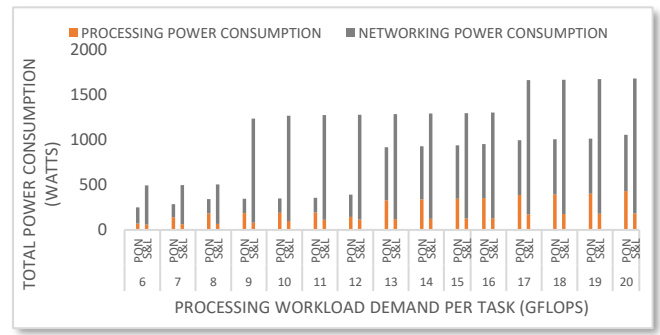

Fig. 14. Power consumption comparison for scenario 3 – case 2.

Fig. 15 shows the average power consumption savings across all scenarios and cases. The results demonstrate that the PON architecture consistently outperforms the S&L. While processing power consumption savings range from favorable (50% in Scenario 1, Case 1) to negative (where S&L outperforms PON in scenario 3), the networking power savings remain remarkably high (59-90%) across all scenarios. This networking efficiency drives substantial total power consumption savings ranging from 39% to 82%.

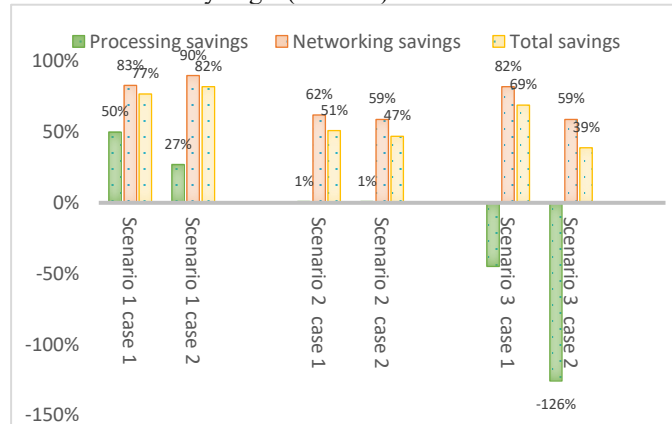

Fig. 15. Average power savings of the PON architecture against S&L architecture across all scenarios.

Fig. 16 presents the power consumption savings achieved by the PON-based fog/cloud architecture compared to cloud processing. Despite processing at the cloud being energy efficient per bit, the idle power overhead negates this advantage. In Scenario1, processing power consumption savings of 89% and 98% are achieved in Case 1 and Case 2, respectively. Networking power savings reached 95% and 92% in the same cases resulting in total power savings of 93% and 89%, respectively. Note that Case1 of Scenario 1 includes demands only from 6 to 11 GFLOPs due to the network bottleneck created when multiple users attempt to utilize the same wavelength simultaneously through a single access point. Similarly, Case 1 of both Scenario 2 and Scenario 3 failed to produce viable allocations due to the same network limitations. In Case 2 Scenario 2, the PON architecture achieves 31% processing power savings and 61% networking resulting in a total power saving of 63% compared to cloud-only processing. The higher demand volumes in Scenario 2 required activating fog nodes in the upper layers of the architecture, which consume more power than lower-layer fog nodes, resulting in reduced overall power savings. Similarly, the second case of the third scenario achieved networking and total savings of 77% and 73%, respectively.

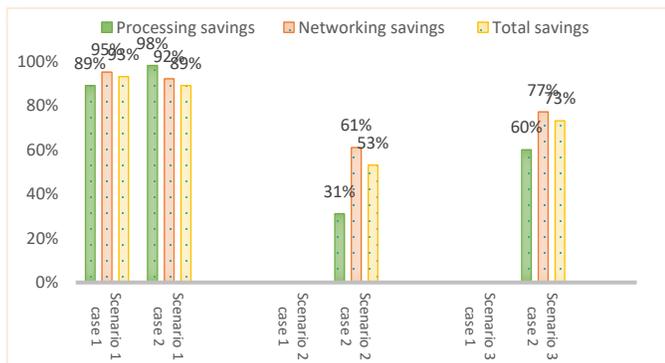

Fig. 16. Average power savings of the PON architecture versus cloud-only processing across all scenarios.

## V Design Enhancement to The Proposed PON Architecture

This section explores different approaches to address the networking limitations in the PON architecture to further enhance its energy efficiency.

### A. Enabling Splitting

We investigate the impact of task splitting on the energy efficiency of the PON architecture. Task splitting enables workloads to be divided into smaller sub-tasks that can be executed simultaneously across multiple fog nodes. This approach can maximize the utilization of available resources.
The MILP model is updated by relaxing Constraint (34), while keeping all other model parameters unchanged. While splitting can offer minimal changes to Scenario 1 where processing already occurred within the building using passive networks, it is expected to produce significant improvements to the other scenarios. To illustrate these effects, we evaluate Case 1 of Scenario 2. Fig. 17 compares the optimal processing placement with and without splitting. For demands 6-12 GFLOPS, splitting produced only minor allocation differences, primarily showing a preference for fully utilizing RFSs before engaging user devices. The most notable impacts occur beyond 10 GFOPs, where splitting consistently prioritize fully packing active resources before activating additional nodes. At 13 and 14 GFLOPs, splitting allowed processing within the building rather than activating processing nodes at higher network layers. Similarly at 15 and 16 GFLOPs, splitting facilitated the utilization of BFS and user devices instead of the CFS. Furthermore, task splitting enabled handling all demands from 17 GFLOPs onwards that were rejected due to network bottlenecks. This is because task splitting allows the model to fully utilize in-building resources while distributing the remaining demands to external nodes.

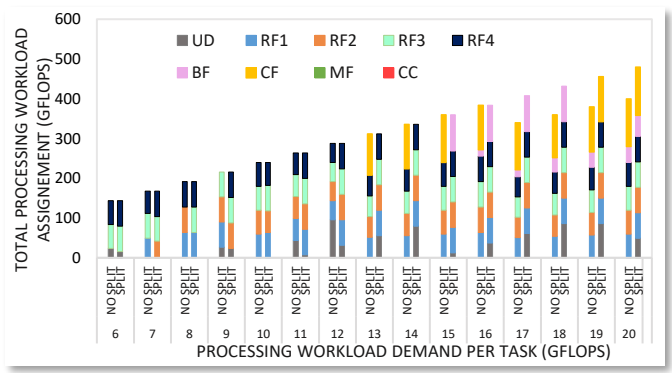

Fig. 17. Optimal processing placement in the PON architecture with and without splitting for scenario 2 – case 1.

Fig. 18 compares the power consumption with and without task splitting for demands ranging from 6-16 GFLOPs. As shown in Fig. 18a, task splitting enabled average processing power consumption savings of up to 37% by utilizing fog nodes closer to users. Additionally, the ability to allocate most of the demands within the building for demands beyond 13 GFLOPs, significantly reduced networking power consumption, enabling average network savings of up to 64%, as presented in Fig. 18b. Fig. 18c shows that task splitting achieved an average total power consumption saving of 54%.

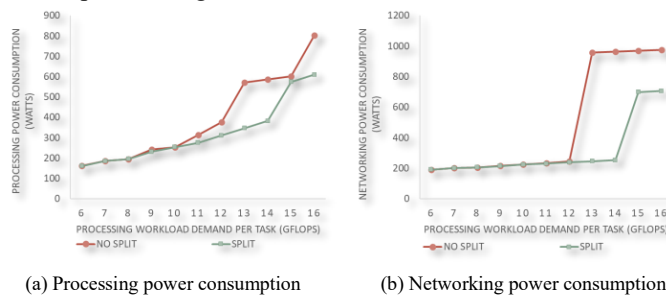

(a) Processing power consumption (b) Networking power consumption

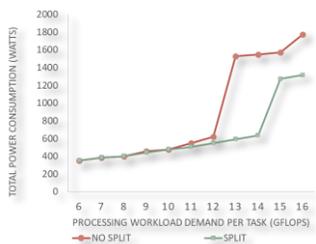

(c) Total power consumption

Fig. 18. Power consumption comparison of the PON architecture with and without splitting for scenario 2 – case 1.

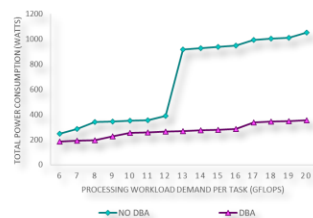

(c) Total power consumption

Fig. 20. Power consumption comparison of the PON architecture with and without DBA.

### B. Dynamic Bandwidth Allocation

This section explores dynamic bandwidth allocation (DBA) as a solution to networking limitations in the PON architecture. Unlike the static allocation where bandwidth is assigned regardless of demand, DBA allocates bandwidth only to active processing resources as needed and the OLT is responsible for optimizing bandwidth based on the ONUs requests. We evaluate the impact of DBA on Case 2 of Scenario 3, where eight user devices in a single room generate processing demands. As shown in Fig. 19, except for the first demand (at 6 GFLOPs), DBA significantly alters processing placement decisions compared to static bandwidth allocation. With DBA, demands are consolidated in fewer RFSs and completely avoiding nodes outside the building for all demands.

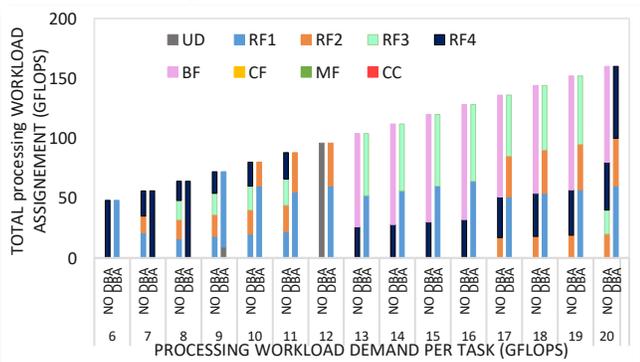

Fig. 19. Optimal processing placement in the PON architecture with and without DBA for scenario 3 – case 2.

Fig. 20 presents the power consumption with and without DBA. As shown in Fig. 20a, DBA significantly reduces processing power consumption by an average of 55% by enabling more efficient resource utilization. Similarly, as presented in Fig. 20b, networking power consumption decreases by an average of 63% through optimized bandwidth management. A total power savings of 59% is achieved compared to the static allocation, as shown in Fig. 20c.

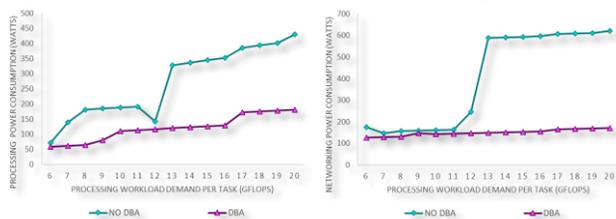

(a) Processing power consumption  (b) Networking power consumption

### C. High Data Rate PON-based Architecture

To address the network bottleneck observed in some of the previous scenarios, particularly in Scenario 3, we investigate the impact of providing sufficient capacity (100 Gbps per wavelength) in the PON architecture (PON1). This enhancement increases the available capacities for user devices and RFSs from 1.1 Gbps to 11 Gbps. This approach aligns with the projected PON network capabilities of 100 Gbps and beyond [21]. We refer to this enhanced PON architecture as PON1 high data rate (PON1-HDR). **Error! Reference source not found.** presents the optimal processing placement results for PON1-HDR and PON1. Since Scenarios 1 and 2 did not exhibit significant network bottlenecks, we focus on Scenario 3 – Case 2, where eight distributed user devices in a single room generate processing demands. The results show that PON1-HDR overcomes the network limitations observed in PON1 by utilizing a single RFS for 6-8 GFLOP instead of multiple user devices or RFSs. At 9 GFLOPs, a single user device is activated in addition to the RFS to support the demands. For 10-16 GFLOPs, two RFS are required in the PON1-HDR while PON1 activates all four. Beyond 13 GFLOPs, the increased link capacity in PON1-HDR eliminates reliance on remote nodes for handling demands.

### D. Devolved AWGR PON-based Architecture

In this section, we propose an improvement to the PON architecture, referred to as devolved PON architecture or PON2, to enable high data rate communication between devices in the same room. As shown in Fig. 21, PON2 maintains the passive nature of the initially proposed architecture but devolves passive connections to the room level rather than between rooms. PON2 incorporates two AWGRs in each room to provide passive connections to the access points within that room to facilitate high data rate connections between the access points and RFS within the same room. PON2 relies exclusively on WDM for traffic conveyance, eliminating the need for splitters and couplers to divide or combine signals. Consequently, each access point is assigned a dedicated wavelength to access other access points and the RFS within the same room. Additionally, PON2 provides passive connections between access points and RFSs in different rooms, based on insights from the original PON architecture (PON1), which demonstrated reliance on the RFSs for processing.

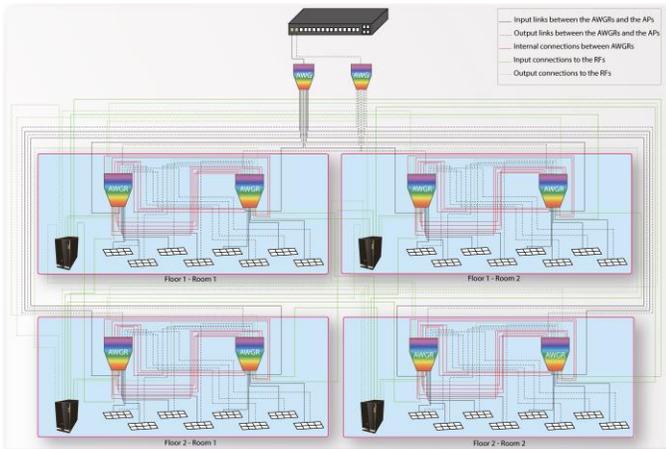

Fig. 21. The devolved PON architecture.

**Error! Reference source not found.** shows the optimal allocation results of PON1 and PON2. PON2 performs similarly to PON1-HDR for 6-8 GFLOPs. However, at 9 GFLOPs, PON2 selects two RFSs instead of a user device due to the absence of direct passive links between user devices within the same building. Beyond 10 GFLOPs, PON2 continues to use only two RFSs like PON1-HDR to avoid offloading to remote nodes beyond 13 GFLOPs.

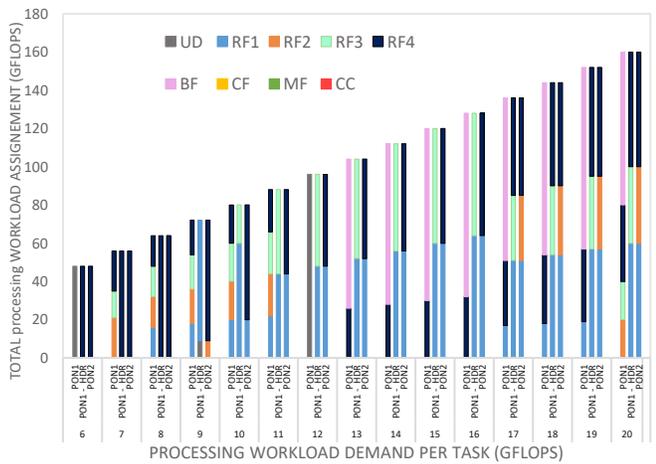

Fig. 22. Optimal processing placement in PON1, PON1-HDR and PON2.

Fig. 23 compares the power consumption of the three designs based on the optimal processing placement. The results indicate that both enhanced designs (PON1-HDR and PON2) significantly reduce power consumption compared to PON1. As shown in Fig. 23a, both designs outperform PON1 by activating fewer processing nodes. The only difference appears at 9 GFLOPs where the activation of RFS in PON2 consumes slightly more power than PON-HDR. Consequently, PON1-HDR achieves 55% average processing power savings compared to PON1, while PON2 achieves 54%. Fig. 23b shows the significant improvement in networking power consumption of both enhanced designs compared to PON1, with PON1-HDR and PON2 being indistinguishable due to their similar passive network utilization, achieving 62% average savings. Both architectures achieve approximately 59% total average savings compared to PON1, as shown in Fig. 23c.

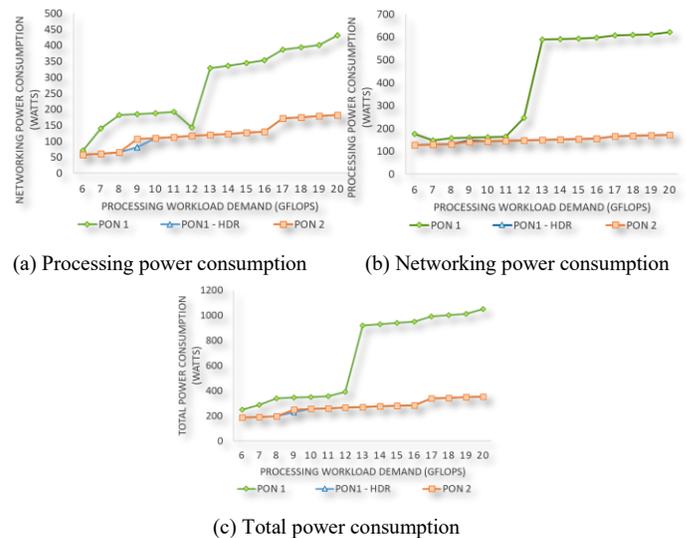

(a) Processing power consumption   (b) Networking power consumption

(c) Total power consumption

Fig. 23. Power consumption comparison between the three designs.

### E. Multi-Building Architecture Extension

To extend the architecture, we introduce an additional building with a similar room layout and resource configuration as the first building, as shown in Fig. 24. To reduce complexity, each room in the second building includes only four access points and one RFS with a total of five devices in each room. Given a fixed 10 Gbps per wavelength, bandwidth is equally divided among the devices in each room. As a result, each device in the second building receives approximately 2 Gbps, compared to about 1.1 Gbps per device in the first building. The two buildings are interconnected via two dedicated OLT ports. The architecture can be scaled up by activating additional ports to connect more buildings.

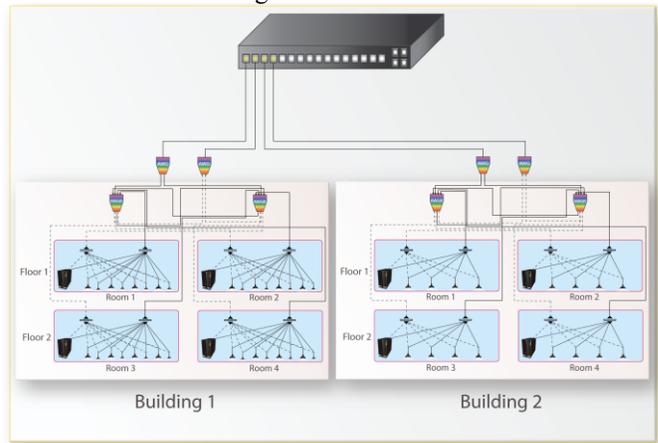

Fig. 24 The extended multi-building PON backhaul network architecture.

Some modifications have been made to the MILP model to incorporate more buildings, including the introduction of new sets and an additional constraint as follows.
Sets:
*BB*        Set of buildings.

$BG_b$     set of the AWGRs in building $b, b \in BB$.
$BIn_{gb}$     Set of the input ports of AWGR $g$ in building $b, g \in BG, b \in BB$.
$BOp_{gb}$     Set of the output ports of the AWGR $g$ in building $b, g \in BG, b \in BB$.

$$\sum_{j \in BIn_{gb}} \lambda_{ijw}^{ud} \leq 0, \quad (50)$$

$\forall u \in UD, d \in P, g \in BG,$
$b \in BB, i \in BOp_{gb}, i \in W, u \neq d$.

Constraint (50) enforces the traffic flow to be directed from the input ports to the output ports of the AWGRs in the second building.

We examine the scenario where six distributed user devices in each room of the first building generate demands while the remaining user devices serve as processing nodes. This scenario is selected due to the high processing demand, which requires the activation of remote fog nodes. Additional idle resources from a nearby building are available to support processing demands, including other user devices (i.e., building 2 user devices (B2UD)), and RFSs in all four rooms (i.e., building two RFS1-4 (b2RFS1-4)). Note that neither task splitting nor DBA is considered in this scenario. Fig. 25 compares the optimal processing placement results of the original (PON) and extended (EX-PON) designs. For lower processing demands (6-12 GFLOPs), both models select RFSs and user devices in the first building. Between 13-16 GFLOPs, PON activates all RFSs and the CFS while EX-PON leverages resources from the neighboring building. Although borrowing resources adds OLT power consumption, this approach is more energy-efficient than routing demands to the CFS, which would incur higher processing and networking power consumption. Beyond 17 GFLOPs, once resources in both buildings are exhausted, EX-PON activates the BFS for remaining demands, unlike PON which adds both the BFS and CFS. EX-PON maintains this allocation strategy, while PON shifts entirely to the MFS from 18 GFLOPs.

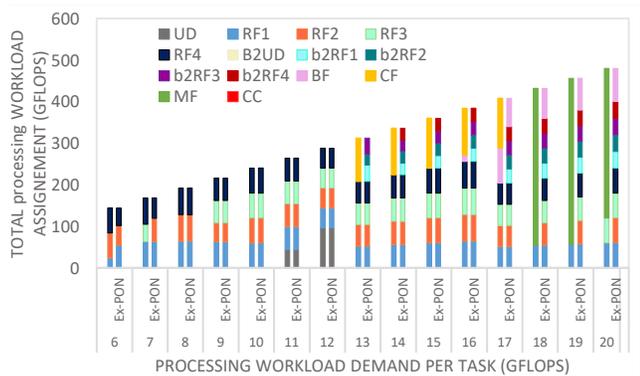

Fig. 25. Optimal processing placement in the original PON and the extended PON architecture.

Fig. 26 illustrates the power consumption of PON and EX-PON architectures. Up to 12 GFLOPs both architectures show identical power consumption trends, as similar resources are activated to handle demands. Although more resources are activated within the two buildings compared to the original architecture, borrowing resources from the nearby building leads to 15% average reduction in processing power consumption, as shown in Fig. 26a. This reduction results from avoiding the high idle processing power consumption associated with resources in higher layers. Additionally, avoiding the higher network layers results in 28% averages savings in networking power consumption, as observed in Fig. 26b. Both processing and networking power consumption contribute to 23% savings in the total power consumption, as seen in Fig. 26c.

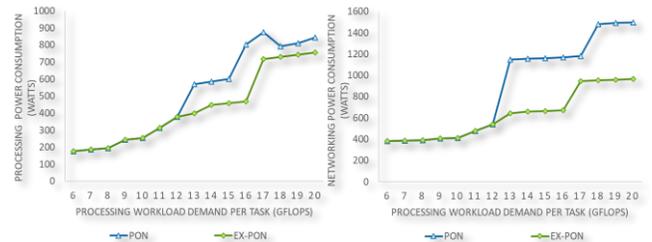

(a) Processing power consumption     (b) Networking power consumption

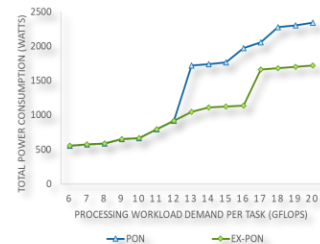

(c) Total power consumption

Fig. 26. Power consumption comparison between the original PON and the extended PON architecture.

## VI. CONCLUSION

This paper demonstrated that VLC is a viable solution for providing connectivity to indoor fog computing environments. We proposed an energy-efficient PON-based backhaul architecture and formulated a MILP model for energy-efficient computing resource allocation. Our evaluation under various workload demands and user distributions revealed significant energy savings, up to 82%, compared to backhaul architectures based on the state-of-the-art spine-and-leaf design. Moreover, the proposed system outperformed centralized cloud processing in terms of energy efficiency, with improvements up to 93%. In addition, we proposed several architectural enhancements including task splitting across multiple processing nodes to improve resources utilization and dynamic bandwidth allocation, increased wavelength bandwidth, and improved connectivity within rooms to alleviate networking bottlenecks. Furthermore, we extended the architecture to connect multi-building to enable borrowing processing resources from neighbouring buildings. This extension significantly enhanced the energy efficiency in high-demand scenarios.

## REFERENCES

[1] Paul Cave, "The Future of Fibre - The Rise of PON ," 2019. Accessed: Sep. 07, 2023. [Online]. Available: https://excel-networking.com/uploads/Whitepapers/future_of_fibre_-_the_rise_of_pon.pdf

[2] R. A. Butt, S. M. Idrus, N. Zulkifli, and M. W. Ashraf, "A Survey of Energy Conservation Schemes for Present and Next Generation Passive Optical Networks.," *J. Commun.*, vol. 13, no. 3, pp. 129–138, 2018.

[3] H. Sadat, M. Abaza, A. Mansour, and A. Alfalou, "A survey of NOMA for VLC systems: Research challenges and future trends," *Sensors*, vol. 22, no. 4, p. 1395, 2022.

[4] G. Lu and W. H. Zeng, "Cloud computing survey," *Applied Mechanics and Materials*, vol. 530, pp. 650–661, 2014.

[5] M. Mukherjee, L. Shu, and D. Wang, "Survey of fog computing: Fundamental, network applications, and research challenges," *IEEE Communications Surveys & Tutorials*, vol. 20, no. 3, pp. 1826–1857, 2018.

[6] F. Jalali, K. Hinton, R. Ayre, T. Alpcan, and R. S. Tucker, "Fog computing may help to save energy in cloud computing," *IEEE Journal on Selected Areas in Communications*, vol. 34, no. 5, pp. 1728–1739, 2016.

[7] J. Bachiega Jr, B. Costa, L. R. Carvalho, M. J. F. Rosa, and A. Araujo, "Computational Resource Allocation in Fog Computing: A Comprehensive Survey," *ACM Comput Surv*, vol. 55, no. 14s, pp. 1–31, 2023.

[8] M. Giordani, M. Polese, M. Mezzavilla, S. Rangan, and M. Zorzi, "Toward 6G networks: Use cases and technologies," *IEEE Communications Magazine*, vol. 58, no. 3, pp. 55–61, 2020.

[9] S. U. Rehman, S. Ullah, P. H. J. Chong, S. Yongchareon, and D. Komosny, "Visible light communication: A system perspective—Overview and challenges," *Sensors*, vol. 19, no. 5, p. 1153, 2019.

[10] J. Song, W. Ding, F. Yang, H. Yang, B. Yu, and H. Zhang, "An indoor broadband broadcasting system based on PLC and VLC," *IEEE Transactions on Broadcasting*, vol. 61, no. 2, pp. 299–308, 2015.

[11] Mark Philip, "Ethernet over Light," The University of British Columbia, 2014.

[12] Y. Wang, N. Chi, Y. Wang, L. Tao, and J. Shi, "Network architecture of a high-speed visible light communication local area network," *IEEE Photonics Technology Letters*, vol. 27, no. 2, pp. 197–200, 2014.

[13] C.-W. Chow, C.-H. Yeh, Y. Liu, C.-W. Hsu, and J.-Y. Sung, "Network architecture of bidirectional visible light communication and passive optical network," *IEEE Photonics J*, vol. 8, no. 3, pp. 1–7, 2016.

[14] H. Kazemi, M. Safari, and H. Haas, "A wireless optical backhaul solution for optical attocell networks," *IEEE Trans Wirel Commun*, vol. 18, no. 2, pp. 807–823, 2018.

[15] A. Hammadi, M. Musa, T. E. H. El-Gorashi, and J. H. Elmirghani, "Resource provisioning for cloud PON AWGR-based data center architecture," in *2016 21st European Conference on Networks and Optical Communications (NOC)*, IEEE, 2016, pp. 178–182.

[16] O. Z. Aletri *et al.*, "Optimum resource allocation in 6G optical wireless communication systems," in *2020 2nd 6G Wireless Summit (6G SUMMIT)*, IEEE, 2020, pp. 1–6.

[17] F. Naqshbandi and R. K. Jha, "TWDM-PON-AN optical backhaul solution for hybrid optical wireless networks," *J Mod Opt*, vol. 63, no. 19, pp. 1899–1916, 2016.

[18] K. C. Okafor, I. E. Achumba, G. A. Chukwudebe, and G. C. Ononiwu, "Leveraging fog computing for scalable IoT datacenter using spine-leaf network topology," *Journal of Electrical and Computer Engineering*, vol. 2017, 2017.

[19] W. B. M. Fadlelmula, S. H. Mohamed, T. E. H. El-Gorashi, and J. M. H. Elmirghani, "Energy Efficient Resource Allocation for Demand Intensive Applications in a VLC Based Fog Architecture," in *2023 23rd International Conference on Transparent Optical Networks (ICTON)*, IEEE, 2023, pp. 1–6.

[20] B. A. Yosuf, M. Musa, T. Elgorashi, and J. Elmirghani, "Energy efficient distributed processing for IoT," *IEEE Access*, vol. 8, pp. 161080–161108, 2020.

[21] N. Feng, M. Ma, Y. Zhang, X. Tan, Z. Li, and S. Li, "Key Technologies for a Beyond-100G Next-Generation Passive Optical Network," in *Photonics*, MDPI, 2023, p. 1128.